\documentclass[10pt,twocolumn,english,preprintnumbers,american,10pt,nofootinbib,amsmath]{revtex4-1}
\usepackage{color}
\usepackage{babel}
\usepackage{graphicx}
\usepackage[unicode=true,pdfusetitle,
 bookmarks=true,bookmarksnumbered=false,bookmarksopen=false,
 breaklinks=false,pdfborder={0 0 1},backref=false,colorlinks=true]
 {hyperref}

\makeatletter

\newcommand{\noun}[1]{\textsc{#1}}

\newcommand{\beq}{\begin{equation}}
\newcommand{\eeq}{\end{equation}}
\newcommand{\bea}{\begin{eqnarray}}
\newcommand{\eea}{\end{eqnarray}}

\usepackage{cancel}
\usepackage[usenames,dvipsnames]{xcolor}

\newcommand{\gev}{\textrm{ GeV}}

\makeatother

\begin{document}
\title{Top quark mass determination from the energy peaks of $b$-jets and
$B$-hadrons at NLO QCD}

\author{Kaustubh Agashe$^{a}$}

\author{Roberto Franceschini$^{b}$}

\author{Doojin Kim$^{c}$}

\author{Markus Schulze$^{b}$}

\affiliation{a) Maryland Center for Fundamental Physics, Department of Physics, University of Maryland, College Park, MD 20742, USA\\
b) Theoretical Physics Department, CERN, Geneva, Switzerland\\
c) Department of Physics, University of Florida, Gainesville, FL 32611, USA}

\preprint{CERN-TH-2016-041, UMD-PP-016-003}

\begin{abstract}
We analyze the energy spectra of {\em single} $b$-jets and $B$-hadrons 
resulting from the production and decay of top quarks within the SM at the LHC
 at the NLO QCD. For both hadrons and jets, we calculate the correlation of the peak of the spectrum with the top quark mass, considering the ``energy-peak'' as an observable to determine the top quark mass. Such a method is motivated by our previous work where we argued that this approach can have reduced sensitivity to the details of the production mechanism of the top quark, whether it is higher-order QCD effects or new physics contributions. 
As part of the NLO improvement over the original proposal, we assess the residual sensitivity of the extracted top quark mass to perturbative effects both in top quark production and decay.
For a  1\% jet energy scale uncertainty (and assuming negligible statistical error), 
the top quark mass can then be extracted using the energy-peak of $b$-jets 
with an error $\pm (1.2 (\hbox{exp}) + 0.6(\hbox{th})) \hbox{  GeV}$.
We note that recently the CMS collaboration reported a top quark mass measurement based on the original proposal (with $b$-jets)
so that our result contributes to a precise evaluation of the associated theory uncertainty.
In view of the dominant jet energy scale uncertainty in the measurement using $b$-jets, we also investigate the extraction of the top quark mass from the energy-peak of the corresponding $B$-hadrons which, in principle, can be measured without 
this uncertainty. The calculation of the $B$-hadron energy spectrum is carried out using fragmentation functions at NLO. The dependence on the fragmentation scale turns out to be the largest theoretical uncertainty in this extraction of top quark mass. Future improvement of the treatment of bottom quark hadronization 
can reduce this uncertainty, rendering methods based on the $B$-hadron energy-peak competitive for the top quark mass measurement.
\end{abstract}
\maketitle

\section{Introduction}

The top quark is a most peculiar particle of the Standard Model (SM), 
%
%
as it experiences both the strongest forces of the SM, {\it i.e.} the strong interaction with gluons and the Yukawa interaction with the Higgs boson, the latter also making it heaviest particle of the SM.
These features make the top-Higgs sector 
a cornerstone of the construction of the SM and of most models of physics beyond the SM, especially those motivated by addressing the Planck-weak scale hierarchy problem. 

A complete knowledge of the top-Higgs sector is therefore a key step for a full exploration of SM physics and for testing models of new physics. 
In fact, the possibility that the SM holds at least as a consistent theory up to the scale of quantum gravity (or Planck scale) requires the Higgs potential to not develop instabilities below this scale, which can be triggered by too strong a top Yukawa interaction, that is too heavy a top quark~\cite{Cabibbo:1979ay}.
For such vacuum stability to hold, the measurements of the Higgs mass require the top quark to be lighter than a value that is within a few GeV from the current top quark mass world combination~\cite{ATLAS:2014fj}\cite{Buttazzo:2013zr,Degrassi:2012kx,Andreassen:2014gha}.
Therefore, a precise measurement of the mass of the top quark is mandatory to discuss the validity of the SM
up to the highest energies. 
Furthermore, the top quark mass is a key parameter for testing the consistency of the SM through precision measurements of the electroweak sector~\cite{Baak:2013yu}.
Additional motivation to measure the top quark mass with high accuracy arises in the context of models of new physics. 
For instance, the prediction of the Higgs boson mass in supersymmetric models has a strong fourth power dependence on the top quark mass and the uncertainty on this mass ends up being a dominant one ~\cite{Djouadi:2008yq,Pardo-Vega:2015oz,Hahn:2014lr,Ellis:1990nz,Barbieri:1990ja}.
%
%
Finally, new types of searches for beyond SM physics in top-like final state can be performed exploiting the precision knowledge of the top quark mass and of its production and decay mechanisms in QCD~\cite{Czakon:2014qf,Czakon:2011fe}\cite{ATLAS-Collaboration:2014dp,ATLAS-Collaboration:2015xw}\cite{Han:2012ve,Cho:2014yma,Ferretti:2015dea,CMS-Collaboration:2016qp,Franceschini:2016tbl}.

All the above considerations give tremendous motivation to measure  the top quark mass accurately,
which is therefore a subject of intense activity at the Large Hadron Collider (LHC) (and was at the TeVatron). 
The present measurements~\cite{TOPLHCWGweb,CMS-Collaboration:2015qy} reach {\em experimental} precision 
%
%
at or below 1 GeV.
%
 %
 In order to support this achieved 
 %
 %
 precision, 
 it is clear that the theory description for hadronic scatterings needs to be accurate.
 The point is that at this level of precision a variety of subtle theoretical effects become relevant, \emph{e.g.} the effects of extra radiation, effects from formation of hadrons, off-shell effects, and more. 
 Some of these effects cannot be computed from first principles, hence require some modelling.
Therefore, it is important to come up with stress-tests for the results of the top quark mass measurement.

Fortunately, there is a way out as follows. 
The large top quark sample collected at Run I, and even more the anticipated data set for Run II, allow us to employ 
%
%
a {\em variety} of methods to extract the top quark mass from the LHC~\cite{CMS-PAS-TOP-11-027,
CMS-PAS-TOP-12-030,CMS-PAS-TOP-15-002,ATLAS-Collaboration:2015qm,Kharchilava:2000yk,Agashe:2013sw,Alioli:2013mz,Juste:2013sp,Kawabataa:2014osa,Frixione:2014jk,Cortiana:2015hp,Corcella:2015hb}, including methods that require rare final states (\emph{e.g.} \cite{Kharchilava:2000yk}).
Each mass determination will (in general) have {\em different} sensitivity to the several theoretical issues that affect top quark physics (in addition to varying experimental systematics). Thus these methods are all complementary, allowing us to study in depth the uncertainties of each method.
A brief survey of the methods currently available follows.
Methods used especially at the TeVatron, \emph{e.g.} the matrix element method of \cite{D0-Collaboration:2015rm}, 
try to exploit the correlation of the top quark mass with each single component of the four-momenta of  {\em all} particles in the top quark final state. 
By doing so, these methods can yield very precise measurements even in presence of a limited data sample. 
The flip side of this approach is that they rely crucially on the theoretical description of the  top quark production and decay, thus being sensitive to a number of possible inaccuracies therein (both the perturbative and the non-perturbative aspects of them).

In order to minimize this sensitivity, several proposals have been made to measure the top quark mass from a well-defined kinematical observable~\cite{Agashe:2013kr}.  
Of course, even in this approach,
in order to extract a measured value of the mass,  we need to compute from theory the correlation between the top quark mass and each of the proposed observables. 
Such a procedure inevitably ties the measurement to our theory description and still exposes it to possible mis-modelling of hadronic collisions at the LHC. 
However, the crucial difference is that in 
relating the top quark mass to a single kinematical observable, one tends to be {\em less} susceptible in this regard
(as follows). 
First, intrinsically the used observable is clearly more inclusive (with respect to {\em other} properties of the events) than the use of the full matrix element as done in the methods mentioned above (i.e. similar to Ref.~\cite{D0-Collaboration:2015rm}). 
Furthermore, this property implies that for some observables 
there is evidently protection from inaccuracies in describing {\em certain} effects. 
For instance, when using purely leptonic observables one can argue that possible uncertainties in the theory description of the formation of hadrons are largely irrelevant. 
Finally, in the computation of the correlation between a single kinematic observable and the top quark mass, it is possible to estimate the theory uncertainties rather straightforwardly, $e.g.$ from scale variations in a perturbative computation. 
Such an error estimation procedure, while still not being perfect, is under better control (and scrutiny) than the one that can be performed on methods similar to Ref.~\cite{D0-Collaboration:2015rm}.

Of course,
there is the possibility of new physics contribution in top quark production which is {\em not} taken into account by both the above classes of methods, since it is assumed therein that the top quarks are produced only by SM interactions.
On the other hand, one can use methods based on Lorentz-invariant observables such as the distribution of invariant mass of decay products or endpoints of transverse mass, which would not be sensitive to new physics, at least for new physics contributing to the top quark sample via the production of on-shell top quarks. 
Thus far, only the CMS collaboration had performed such an experimental analysis~\cite{CMS-PAS-TOP-11-027}.

In summary then, for all the above-mentioned reasons, it is prospected that at the end of the LHC program, the implementation of several of those methods (and any future ones) for the top quark mass measurement will provide a powerful consistency check and thus enable a {\em global} determination of the top quark mass at the LHC with an uncertainty evaluated with much higher confidence than that in any single method.

In light of the above discussion, in this paper we systematically develop a new method for the precision determination of the top quark mass  with higher order corrections taken into account, based on the energy distribution of  $b$-jets and $B$-hadrons resulting from its decay, which promises to add significant information to the combination of methods presently available. 
In particular, we will show in detail that the present method has reduced sensitivity to mis-modelling of the production mechanism of the top quark at the LHC, including possible production of top quarks from new physics.
%
%
%

%
%
Our method refines and extends the original proposal in Ref.~\cite{Agashe:2013sw}:
we contextualize our contribution by first recapitulating this earlier work.
It is very well-known that, in the rest-frame of the top quark undergoing a (strictly) two-body decay into a bottom quark and $W$ boson, the energy of the bottom quark is single-valued:
\begin{equation}
\hat{E}_{b}^{\hbox{\small (LO)}}=\frac{m_{t}^{2}-m_{W}^{2}}{2m_{t}}\,, \label{eq:EbLo}
\end{equation}
which is nothing but a simple function of the masses involved.\footnote{Here, the bottom quark mass is neglected for simplicity.}
However, in the laboratory frame at a hadron collider, the top quark is produced with a non-zero boost, that too 
in general different in one event and the next.
Thus, the above $\delta$-function of the bottom quark energy in the rest-frame gets smeared into a 
non-trivial observed energy spectrum, which varies with the boost distribution of the top quark (as expected due to energy being a Lorentz-variant quantity).  
In turn, the boost distribution of the top quark depends on details of its production, including the matrix element involved and external parameters such as the collider energy and relevant parton distributions.

Remarkably, it was shown in Ref.~\cite{Agashe:2013sw} that, assuming the top quark is produced unpolarized, the location of the maximum of the inclusive bottom quark energy distribution (dubbed the ``energy-peak'') remains the same as the (single) value in the rest-frame. 
To stress the significance of such a result, we underline that the statement is valid irrespective of the boost distribution of the top quark, that is to say the energy-{\em peak} is ``invariant'' in this sense.

The upshot of the above argument is that measuring this energy-peak can (in principle) afford us the opportunity
for a determination of the top quark mass which can be rather insensitive to uncertainties in top quark production, whether arising from higher order (as yet uncalculated) QCD effects within the SM or any possible
new physics contribution beyond it. 
In other words, energy-{\em peak} methods, being based on an invariance of the decay, have the potential to enjoy several of the benefits of mass measurement which are traditionally reserved for manifestly Lorentz invariant quantities mentioned earlier, 
even though naively this looks implausible, given that energy by itself is {\em not} invariant.
Moreover, energy-peak methods, being based on {\em single}-particle measurements, might not share some of the drawbacks of those other methods, for example, combinatorial ambiguities (for both invariant and transverse mass) or reliance on the use of the missing transverse momentum (for the latter).
Because of these possible advantages offered by energy-peak methods, we expect that a top quark mass determination following them will be complementary to existing ones.

Of course, as with other methods, one has to consider robustness of the energy-peak to going beyond the above framework.
First of all, the bottom quark from the top quark decay manifests in the detector as a $b$-jet containing a $B$-hadron.
Therefore, one has to correlate the above-discussed energy-peak of the bottom {\em quark} to that of what is actually observed.
For example, we expect that hard radiation in the decay of the top quark, {\it i.e.} $t \to b W g$, when not reclustered in the $b$-jets that we analyze will lead to deviations of the $b$-{\em jet} energy-peak from Eq.~\eqref{eq:EbLo}, hence the superscript ``LO'' there.
This effect is analogous to the softening of the di-lepton invariant mass spectrum from decays of the $Z$ boson due to the escaped hard radiation, $i.e.$ $Z\to \l^{+}\l^{-}\gamma$, and was already mentioned in our work Ref.~\cite{Agashe:2013sw}.
In addition, as discussed already in Ref.~\cite{Agashe:2013sw}, in a realistic situation we cannot deal with a fully inclusive spectrum, due to cuts needed to remove backgrounds, leading to loss of a strict invariance. 
With these considerations, the goal of the present paper is to study in detail the above higher-order effects on the energy-peak methods for measuring the top quark mass. As we will see quantitatively, this improvement on the earlier result of Ref.~\cite{Agashe:2013sw} is necessary in order to employ this method for a precision top quark mass measurement at the LHC.
In particular, assuming pure QCD production, we first present a calculation at the next-to-leading order (NLO) 
in $\alpha_s$ of the energy distribution of the {\it single} $b$-{\em jets} from the top quark production and decay at the LHC~\footnote{A related computation
already appears in \cite{Biswas:2010vq}, where the distribution the of {\em sum} of the two $b$-jet energies at NLO was shown.}, 
thus computing the 
{\em shift} in the energy-peak from Eq.~(\ref{eq:EbLo})  
at this order (originating primarily from the above-mentioned radiation in the decay).
Note that the size of this shift is estimated to be of order $\alpha_s / \pi$, that is {\em few} percent (i.e.
%
%
several GeV) level, relative to the LO prediction. 
This effect constitutes  the main source of theory error in our earlier work Ref.~\cite{Agashe:2013sw}: it is then 
critical to calculate this NLO effect, as we do in this work, to reach the $\sim 1$ GeV  uncertainty of current methods used to measure the top quark mass.
With the NLO result at hand, we assess the size of the {\em residual} sensitivity to QCD corrections in production and decay of the top quark.
Our study indicates that a perturbative theory uncertainty of about 1 GeV can be achieved, establishing a precision attainable with the energy-peak methods which is comparable to that of other existing methods.

We reiterate
%
%
here that, thanks to the above-mentioned peak invariance, the sensitivity of the measured top quark mass to any uncertainties in the understanding of the {\em production} of top quarks, including that from higher-order QCD corrections, is {\em reduced} for this method (as compared to other methods).
This ``screening'' happens because the actual measurement carried out at the experiments is not far from the ideal case for which  the invariance result strictly applies. 
In other words, the experimental conditions allow us to go quite close to the measurement of the inclusive energy spectrum of a bottom quark, that is a jet which contains all the radiation of the parent quark and no pollution from other radiation, without requiring too hard cuts to remove backgrounds. 
This implies that the effects from mis-modelling the relevant production are felt only through small perturbations on the original invariance, that are weighted by a factor of order $ \alpha_s / \pi$ for effects from hard radiation from the bottom quark or a similarly small factor for the non-inclusiveness of the spectrum due to selection cuts.
Extending the above discussion to possible new physics contribution to production of top quark, we see that the sensitivity of the energy-peak to such effects is similarly suppressed, because also in this case it is weighted by the above small factors (or by the polarization fraction and velocity of the top quarks, in case they have a preferred polarization state).
This should be contrasted with the methods that assume the SM matrix element from the very beginning, which might (in principle) face the {\em full} impact of this uncertainty.
All in all, the above discussion shows that our work is bringing important new elements for the current precision measurement of the top quark mass from the LHC~\cite{Juste:2013sp,CMS-PAS-FTR-13-017} and similarly for the future runs.

As a matter of fact, following our earlier proposal in Ref.~\cite{Agashe:2013sw}, the CMS collaboration in Ref.~\cite{CMS-PAS-TOP-15-002} has already measured the top quark mass from the location of the peak in the $b$-jet energy spectrum at Run~I, although without the full NLO corrections in the theory used to interpret the measurement. 
The results of the present work will thus be crucial for a precise evaluation of the theory uncertainty on this  determination of the top quark mass during the next iteration of such measurement.

As is true for many of the methods presently formulated, the measurement of the top quark mass from the $b$-jet energy spectrum will suffer from the experimental  jet energy scale uncertainty. 
Attempting to surpass this obstacle for future high-precision measurements, we explore the possibility to extract the top quark mass using the 
%
%
energy spectrum of the $B$-hadrons contained in a $b$-jet. 
This spectrum in principle can be be measured using only the
tracker sub-detector,
hence liberating the measurement from the jet energy scale uncertainty, for instance, exploiting the decays of $B$-hadrons into charged hadrons, such as $B^{+}\to K^{+}\pi^{+}\pi^{-}$ or any of the $J/\psi$ modes used in Ref.~\cite{Kharchilava:2000yk}.
Anticipating the exploitation of this exciting possibility at the time when the LHC data set will be large enough to study these rare decays, we perform the calculation of the energy-{\em peak} of the $B$-hadrons 
using fragmentation functions accurate at NLO in QCD. We study the correlation 
of this observable with the top quark mass and the theory error on a possible measurement of the top quark mass
using it. 
We find that the present {\em theoretical} description of the formation of $B$-hadrons gives errors slightly larger than those of the  existing methods for top quark mass measurement, but the theory improvements that might be available by the time the LHC will be ready to study these rare decays might lead to a competitive measurement.


This paper is organized as follows: in Section~\ref{sec:Energy-peaks} we recall relevant aspects of the physics of energy distributions (in particular, their peaks) from two-body decays presented in Ref.~\cite{Agashe:2013sw} and we describe the challenges and expected results of the extension to NLO of previous LO results. 
In Section ~\ref{sec:B-jet-Results} we present our results for the 14 TeV LHC for the measurement of the top quark mass from the analysis of energy-peaks of (single) $b$-tagged jets from top quark pair production in the $e\mu$ final state for the two $W$ bosons.
In Section~\ref{sec:B-hadrons-results} we present analogous results based on calculations of the single $B$-hadron energy spectrum obtained from the $pp\to b\bar{b} \ell \bar{\ell} \nu \bar{\nu}$ process supplemented with fragmentation function at NLO QCD~\cite{Biswas:2010vq}. 
We present our conclusions and outlook in Section~\ref{sec:Conclusions}.

\section{Energy spectra and the top quark mass\label{sec:Energy-peaks}}
We begin with a brief review of 
the energy spectra of massless two-body decay products, assuming that  
a sample of decaying particles has no preferred polarization state: it turns out that these spectra 
enjoy special properties within relativistic kinematics.
A good example to accommodate this situation is the top quark (produced at the LHC) decaying into a bottom quark, which is approximately massless compared to its typical energy scale, and an (on-shell) $W$ gauge boson.
Furthermore, if pair-produced, the top quark is unpolarized since it is produced predominantly via QCD. 
For such a top quark decay, we then obtain a $b$ quark energy distribution peaking at the value of Eq.~\eqref{eq:EbLo}~\cite{Agashe:2013sw}.\footnote{See also \cite{1971NASSP.249.....S} for analogous arguments valid for $\pi^{0}\to\gamma \gamma$ decay.}
The superscript ``LO'' in Eq.~\eqref{eq:EbLo}  stresses that this relation holds for a strictly two-body decay of the top quark. 
Eq.~\eqref{eq:EbLo} implies that a measurement of the energy peak together with the known value of the $W$ boson mass enables us to determine the mass of the top quark.

Strictly speaking, this result is valid for a fully inclusive sample of $b$ quarks from the top quark decay. 
Therefore, the peak position predicted in Eq.~\eqref{eq:EbLo} will shift if one restricts the  analysis only to a subset of the $b$ quarks ({\it e.g.} the ones detected in the most sensitive region of the experiment). 
Furthermore, it is inevitable that a fraction of top quark decays contain extra hard emissions, in particular, final state radiation resulting in three-body decays of the top quark which would lead to a violation of the relation in~\eqref{eq:EbLo}~\cite{Agashe:2012ij}.
On top of these issues, the actual observable is {\it not} an isolated $b$ quark {\it but} a $b$-tagged jet reconstructed from detected hadrons, causing additional subtleties. 
Indeed, Ref.~\cite{Agashe:2013sw} performed an analysis to investigate the potential effects from the above-mentioned issues and found that they are not a major hurdle in determining the top quark mass at leading order accuracy with the $b$-jet energy-peak technique. 
In this work we shall re-examine the effect of taking a non-inclusive sample of $b$-jets (as unavoidable for experimental reasons) and that of extra radiation, with the goal of $i)$ making a statement on those effects by treating the first emission in QCD (correcting both production and decay) without any approximations and $ii)$ estimating the potential impact of multiple emissions in the parton shower approximation.

In next two sections, we study the prediction of perturbative QCD and its theory uncertainty upon the {\it relation between the top quark mass and a physical observable}.
For this purpose, we consider the maximum in the energy spectrum of $b$-jets (Section~\ref{sec:B-jet-Results}) and $B$-hadrons (Section~\ref{sec:B-hadrons-results}) subject to typical selection cuts. 
We begin with the perturbative QCD prediction for the differential energy distribution 
\bea
\frac{d\sigma}{dE_{b-tag}}\,\,\mbox{ for the process \,\,}pp\to t\bar{t}\to b\bar{b}e^{-}\mu^{+}+\nu\bar{\nu}\,,\label{eq:process}
\eea
where $E_{b-tag}$ denotes the energy of any $b$-tagged jets (tagged hadrons) in the given event. 
From the shape of this distribution we determine its maximum point,  which we henceforth denote by $\hat{E}_{b-tag}$. 
Since typical energy spectra are broadly distributed, in practice, we obtain the maximum point by fitting $\frac{d\sigma}{dE_{b-tag}}$ to a function.
The detailed fitting procedure will be given when each of the specific applications are discussed later.

As obvious from Eq.~\eqref{eq:EbLo}, the LO perturbative QCD prediction for $\hat{E}_{b-tag}$ has a simple dependence on the top quark mass.
In general, in the vicinity of any fixed top quark mass value $\bar{m}_t$, the functional dependence of $\hat{E}_{b-tag}$ over $m_t$ can be linearized:\begin{equation}
\hat{E}_{b-tag}=\rho_{b} (m_{t}-\bar{m}_t)+\textrm{const}+\mathcal{O}\left((m_t-\bar{m}_t)^2\right)\,,\label{eq:EbGenericRelation}
\end{equation}
where the coefficient $\rho_{b}$ encodes the correlation between the physical observable to be measured in experiments (the peak of the energy distribution) and the SM parameter we intend to measure (the top quark mass). At the lowest order in perturbation theory and for fully inclusive spectra, the prediction in Eq.~\eqref{eq:EbLo} gives $\rho_{b}\simeq0.6$ and $\textrm{const}\simeq 68\gev$ for $\bar{m}_{t}=173\gev$. 
Beyond the lowest order in perturbation theory, the prediction for the right hand side of Eq.~\eqref{eq:EbGenericRelation} depends on details of the calculation such as the order in perturbation theory and the choices for the renormalization and factorization scales.
The difference between the predictions by equivalent theory calculations ($e.g.$ varying scale choices or other unphysical parameters) can provide a measure of the theory uncertainty for the relevant prediction.  
To assess this uncertainty we perform several variations in our calculations determining the prediction in Eq.~\eqref{eq:EbGenericRelation}. 
We give full detail our assessment scheme for each of the mass measurement applications later.
Finally, in order to use the relation in Eq.~\eqref{eq:EbGenericRelation} beyond the lowest order in perturbation theory, we need to specify what is the $m_{t}$ in its right hand side. 
In what follows, we shall take the fixed order calculation in the pole mass scheme, and therefore, unless noted otherwise, $m_{t}$ throughout this work is the top quark pole mass. 

\begin{figure}
\begin{centering}
\includegraphics[width=1\columnwidth]{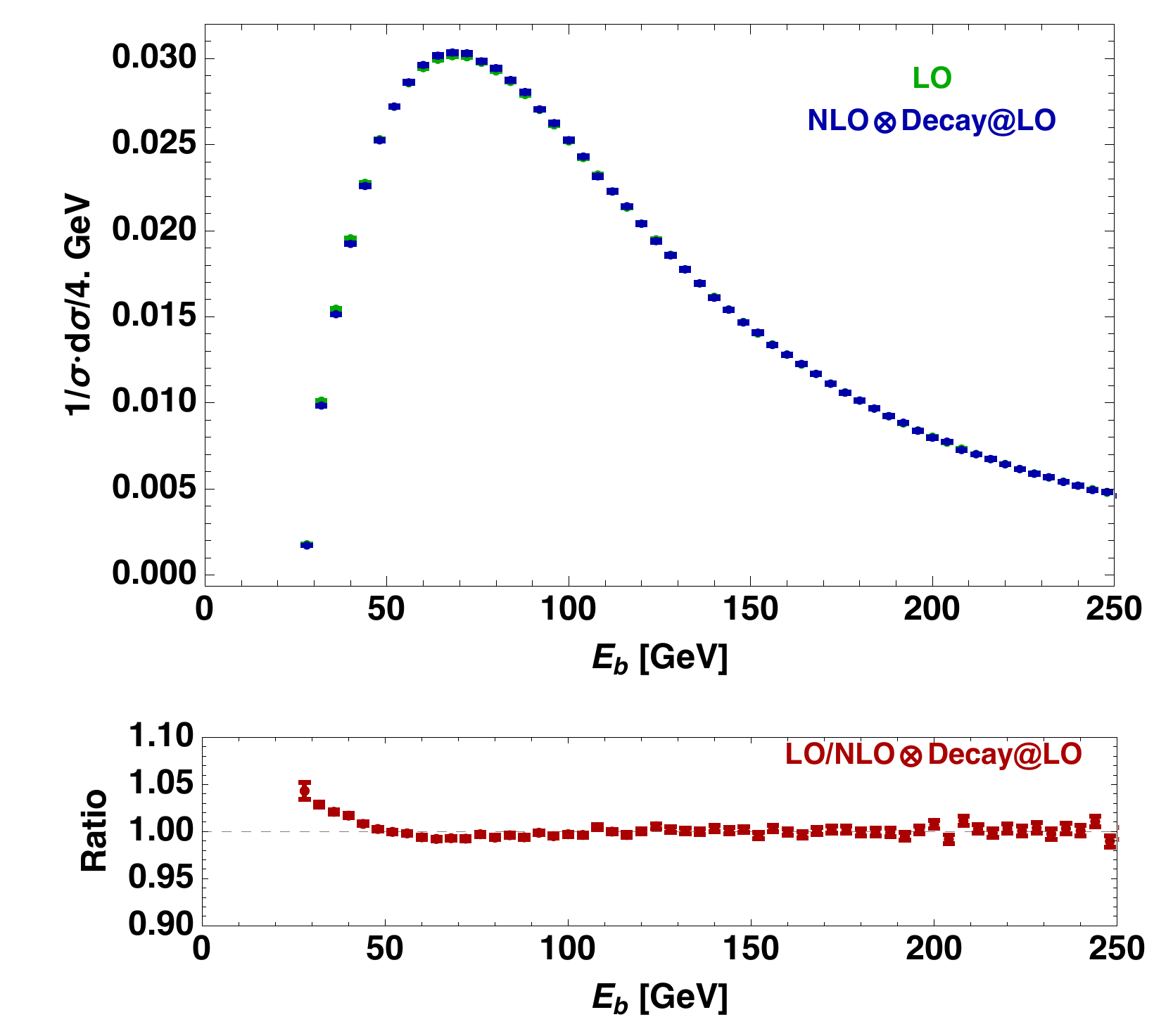}
\par\end{centering}

\caption{\label{fig:LOvsPNLO}
Prediction of energy spectrum under the selection criteria in Eq.~\eqref{eq:cut1} with production of the $t\bar{t}$ at LO (green, process Eq.~(\ref{LOprocess})) and NLO (blue, process Eq.~(\ref{pNLOprocess})) for $\mu_{F}=\mu_{R}=m_{t}=171\hbox{ GeV}$. The bottom panel gives the ratio of the two spectra and highlights the tendency of the NLO production correction to harden the spectrum as expected  from extra radiation affecting the $b$-jet energy through clustering.}
\end{figure}

Before describing the actual calculations and their results, it is worth discussing the new effects that arise once we go beyond the leading order in perturbation theory. 
Compared to the lowest order description, the addition of extra radiation to the production and decay of the top quark induces several modifications in the energy distribution of the decay products of the top quark decay, some of which also affect our observable $\hat{E}_{b-tag}$. 
As we present our calculations with the production and decay of top quarks factorized, we can discuss separately the potential effects in the production and decay of top quarks.
Suppose that the LO process is represented by 
\bea
pp \rightarrow t\bar{t} \otimes t \rightarrow bW\,,\label{LOprocess}
\eea
where we symbolize the factorization of the top quark decay by $\otimes$ and leave implicitly the charge conjugate decay of $\bar{t}$ for notational simplicity.
NLO corrections to the production mechanism modify the above-given LO expression into
\bea
pp\rightarrow t\bar{t}+0 \hbox{ or }1 \hbox{ jets}\otimes t \rightarrow bW\,. \label{pNLOprocess}
\eea
The extra radiation accompanying the production
of the $t\bar{t}$ pair does not affect the decay of the top quarks \emph{per se}. 
It merely changes the boost of the $t\bar{t}$ system, hence the boost from the rest frame of each top quark to the laboratory frame.
Therefore, one can expect that the existence of extra radiation in production will affect the overall shape of $d\sigma/dE_{b-tag}$. 
However, due to the fact that the energy-peak is insensitive to the details of the boost distribution of the decaying particle, any extra radiation at the production level {\it does not alter the LO relation between our observables and the top quark mass}.
This implies that, very strikingly, Eq.~\eqref{eq:EbLo} is still valid for inclusive bottom quark energy spectra computed with NLO corrections to the production mechanism and LO descriptions of the decay. 
We point out that this production-level radiation can still make an effect on our observable in the reconstructed $b$-jet energy spectra, because  the jet clustering can cluster in the $b$-jet radiation that was not originated by the $b$ quark. 
Clearly, this effect becomes increasingly significant as the size of the jets grows. We have checked explicitly that it grows proportionally to the jet area, that is to say  $\hat{E}_{b-tag} = \textrm{const} + \textrm{coeff} \cdot R^{2}$. Furthermore, we also found that for inclusive $b$-jet spectra $\textrm{const}=\hat{E}_{b}^{LO}$ of Eq.~\eqref{eq:EbLo}, hence the effect disappears for infinitesimally small jets, as implied by the insensitivity of the energy peak to the boost distribution of the top discussed in  Ref.~\cite{Agashe:2013sw}.
The effect for the realistic jet size $R=0.5$, which is employed throughout our analysis, is fully taken into account in our top quark mass determination. An example of the hardening of the spectrum caused by clustering effects is shown in Figure~\ref{fig:LOvsPNLO}
{for $\mu_{F}=\mu_{R}=m_{t}=171\hbox{ GeV}$ and  selection cuts given in Eq.~\eqref{eq:cut1}.
The bottom panel of the figure demonstrates the ratio of the differential energy spectrum from the process Eq.\eqref{LOprocess} to that from the process Eq.~\eqref{pNLOprocess}, from which we clearly see a (slight) hardening effect in the low energy regime.}

\begin{figure}
\begin{centering}
\includegraphics[width=1\columnwidth]{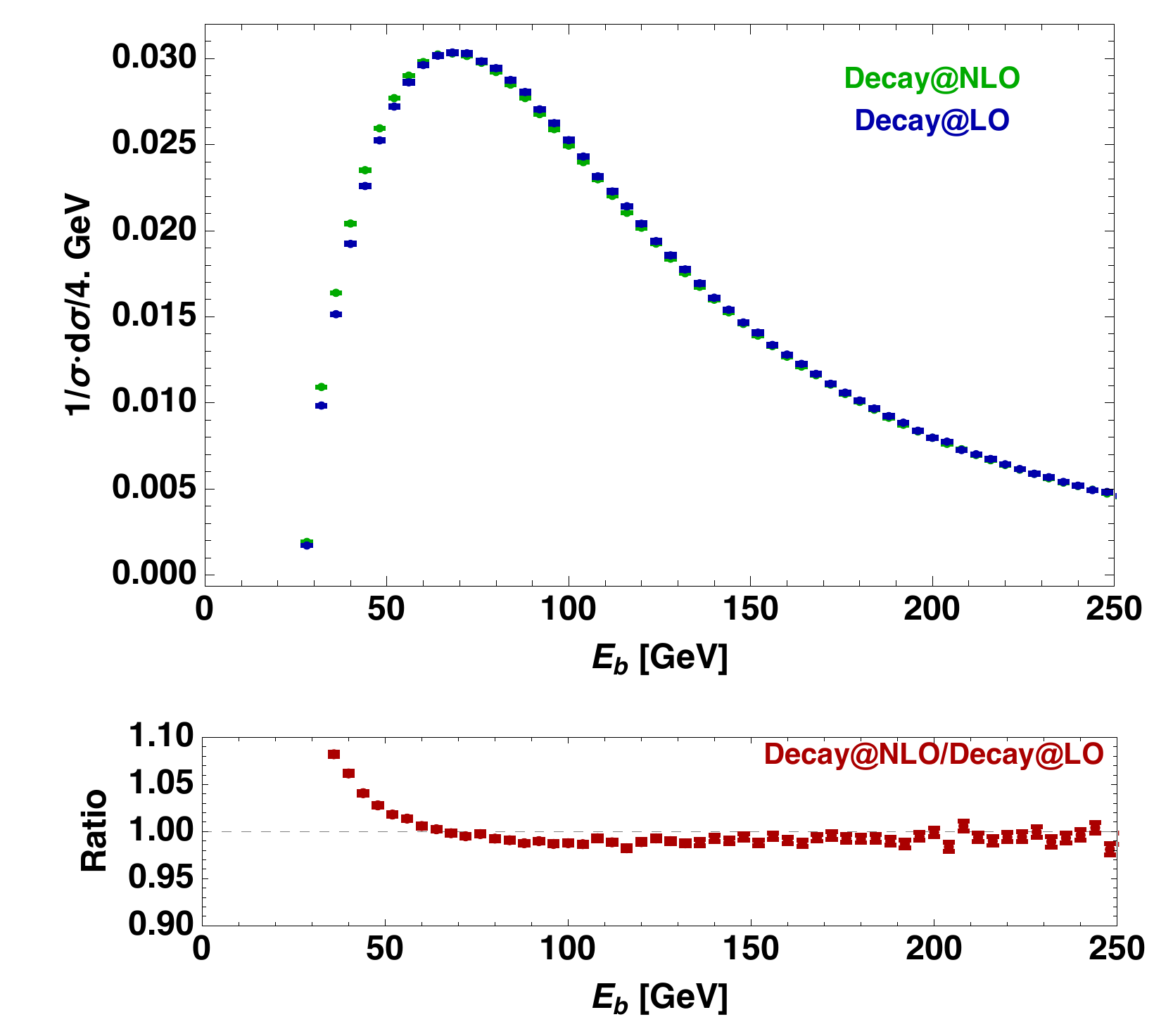}
\par\end{centering}

\caption{\label{fig:DecayLOvsNLO}
Prediction of energy spectrum under the selection criteria in Eq.~\eqref{eq:cut1} with decay at LO (blue, process Eq.~(\ref{pNLOprocess})) and NLO (green, full NLO process Eq.~(\ref{eq:process})) for $\mu_{F}=\mu_{R}=m_{t}=171\hbox{ GeV}$. The bottom panel gives the ratio of the two spectra and highlights the tendency of the NLO decay correction to soften the spectrum as per Eq.~(\ref{eq:NLOshiftJet}).}
\end{figure}

When it comes to the decay NLO, the relevant corrections modify the LO process into 
\bea
pp\to t\bar{t}\otimes t\to bW\,+\mbox{0 or 1 jets}.
\eea
In this case the $b$-tagged object is systematically softer because some fraction of energy is carried away by the extra resolved parton that emerges as a separate jet from the top quark decay. 
Under this circumstance, both the overall shape \emph{and} the peak in $d\sigma/dE_{b-tag}$ are different compared to the leading order case. 
In fact, the decay with an extra jet is a three-body decay for which, as established in great detail in Refs.~\cite{Agashe:2015wj,Agashe:2012ij}, the expected maximum point is lower than that in the LO case, $i.e.$ $\hat{E}_{b-tag}<\hat{E}_{b}^{(LO)}$. 
This \emph{inequality is the only genuine NLO correction that we expect} for this kind of extraction of the top quark mass. 
This remark is very important because it allows us to estimate the higher order corrections of the peak position such that 
\begin{equation}
\frac{\hat{E}_{b-tag}}{\hat{E}_{b}^{(LO)}}\simeq1-|d|\frac{\alpha_{s}}{\pi}\,,\label{eq:NLOshiftJet}
\end{equation}
where $d$ is a coefficient encapsulating the dependence on several parameters ($e.g.$ jet definition, top quark mass, scales used in the calculation) but is not expected to be subject to any accidental enhancement,  for example, from large logarithms or large increases of initial state luminosity, as it can happen in calculating corrections to {\em production}. An example of effect of adding radiation in decay is displayed in Figure~\ref{fig:DecayLOvsNLO},
{for $\mu_{F}=\mu_{R}=m_{t}=171\hbox{ GeV}$ and selection cuts given in Eq.~\eqref{eq:cut1}.
In particular, the bottom panel shows the ratio of the differential energy spectrum from the process \eqref{eq:process} at production and decay NLO to that from the process \eqref{pNLOprocess},}
which manifests well the softening of the spectrum due to NLO corrections to the decay of the top quark.

The case with the $B$-hadron can be understood in an analogous manner to the previous case, given that it emerges from the $b$ quark as the result of $t\rightarrow Wb \rightarrow WB+X$. 
Here the ``$X$" system surrounding the $B$-hadron (i.e. the {\em rest} of the $b$-jet) draws a certain fraction of energy
of the parent bottom quark.
This picture 
%
%
resembles a three-body decay of the top quark, i.e. similarly to $t \rightarrow bW + g$ discussed earlier, 
so that another inequality  $\hat{E}_{B}<\hat{E}_{b}^{(LO)}$ is anticipated for the $B$-hadron case (cf.~Eq.~(\ref{eq:NLOshiftJet})).
In this case, the discrepancy between the $B$-hadron energy-peak and the LO result in~\eqref{eq:EbLo} is governed by the physics of hadronization. 
Note that the strong coupling constant relevant to hadronization is larger than that relevant to extra emission and the associated final state that we investigate is more exclusive than the $b$-jet final state. 
We therefore expect that the corresponding effect in this case will be larger, and possibly converging less fast than that in the $b$-jet case.

\section{$b$-jet Results\label{sec:B-jet-Results}}

We now present the detailed calculations of the energy-peak 
method for top quark mass measurement using $b$-jets.
We compute energy spectra of $b$-tagged jets using\noun{ mcfm~6.8}~\cite{Campbell:2012rt} cross-checked by the code from Ref.~\cite{Melnikov:2009nx,Biswas:2010vq} including real and virtual corrections to the process in Eq.~\eqref{eq:process}. 
The calculation is carried out using the \noun{mstw08nlo} PDFs set~\cite{Martin:2009zm} at the center of mass energy $\sqrt{s}=14\mbox{ TeV}$.
We treat the top quark in the narrow-width approximation and include NLO spin correlations. 
We then set the renormalization scale $\mu_{R}$ and the factorization scale $\mu_{F}$ to be equal for our baseline calculation, and vary them within the range of $[0.5m_{t},2m_{t}]$ to obtain a measure of the theory uncertainty in the calculation. 
Jets are constructed from the final state partons using the anti-$k_{T}$ algorithm~\cite{Cacciari:2008hb} with size parameter $R$ being 0.5. 
An event is kept to fill the $d\sigma/dE_{b-tag}$ distribution if it contains exactly two leptons and satisfies the following set of cuts:
\begin{equation}
p_{T,\ell}>20\mbox{ GeV},\,|\eta_{\ell}|<2.5,\,p_{T,j}>30\mbox{ GeV},\,|\eta_{j}|<2.5\,.\label{eq:cut1}
\end{equation}
Any jet containing at least one $b$ quark as a jet constituent is identified as a $b$-jet. 
We require exactly such two $b$-tagged jets, so that from each event we get two contributions to fill into the $b$-jet energy distribution.

For each choice of scales $\mu_F$ and $\mu_R$ the relevant calculation yields an energy spectrum $d\sigma/dE_{b-tag}$ from which we find the maximum point by fitting it to a fitting function and finding the best fit.
The fit is performed as a least-$\chi^{2}$ fit, and our model function $f(E_{b-tag})$ has the form 
\bea
f(E_{b-tag})\sim\exp\left[-w\left(\frac{\hat{E}_{b-tag}}{E_{b-tag}}+\frac{E_{b-tag}}{\hat{E}_{b-tag}}\right)\right]\,.
\eea
This function was used in our previous LO study \cite{Agashe:2013sw} and found to be suitable to describe the NLO spectrum as well. 
From the fit we obtain the maximum point $\hat{E}_{b-tag}$, a measure of the width of the spectrum parametrized by $w$, and their statistical errors obtained from $\chi^{2}$ variations.

\begin{figure}
\begin{centering}
\includegraphics[width=1\columnwidth]{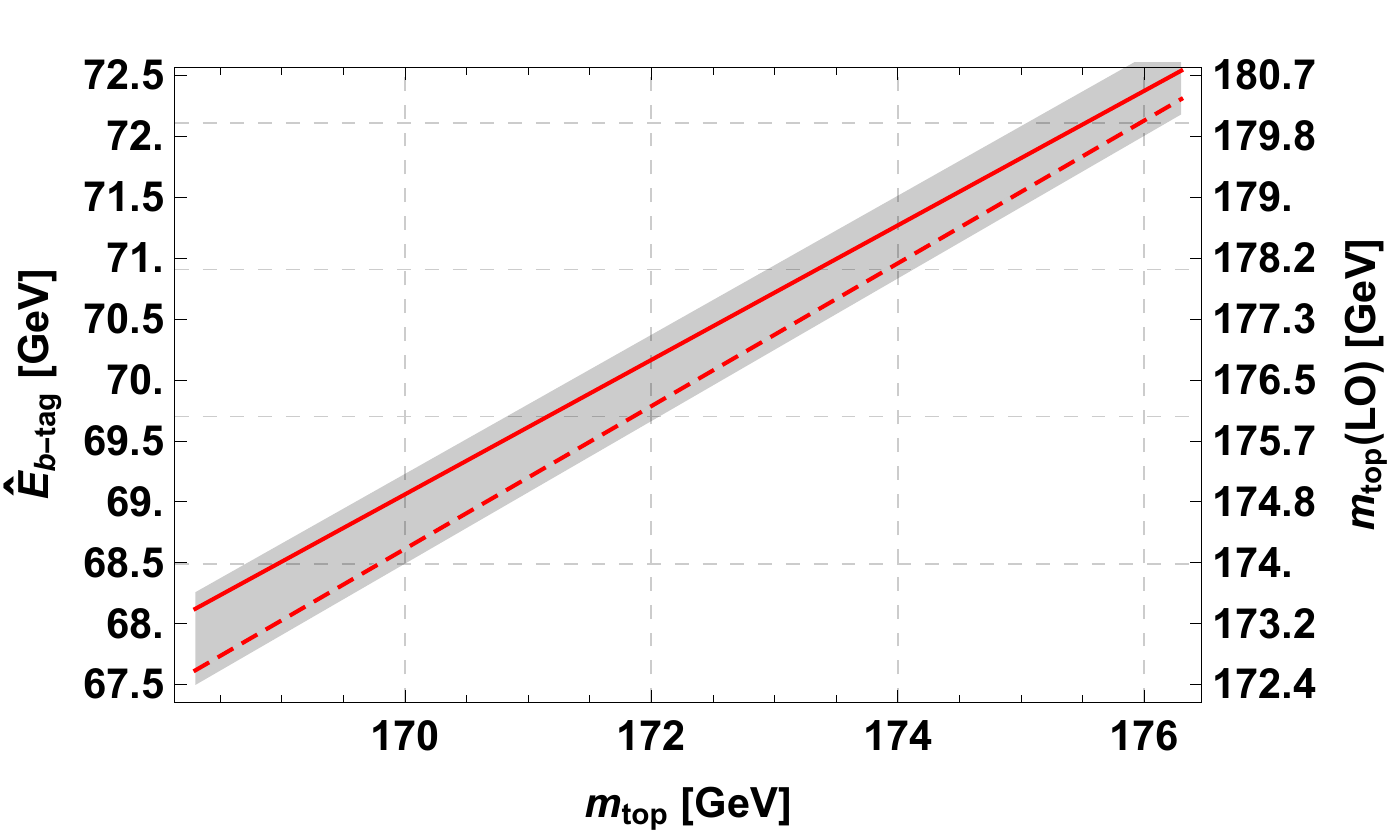}
\par\end{centering}

\caption{\label{fig:Predicted-energy-peak-jet}
Prediction of the fitted energy peak as a function of $m_{t}$, as obtained from fitting the part of the energy spectrum where $45\mbox{ GeV}<E_{b-tag}<160\mbox{ GeV}$ under the selection criteria in Eq.~\eqref{eq:cut1}. 
To ease the reading of the resulting uncertainty on $m_{t}$, along the right vertical axis we give the top quark mass corresponding to the $\hat{E}_{b-tag}$ according to the LO formula in Eq.~\eqref{eq:EbLo}.}
\end{figure}

\begin{figure}
\begin{centering}
\includegraphics[width=1\columnwidth]{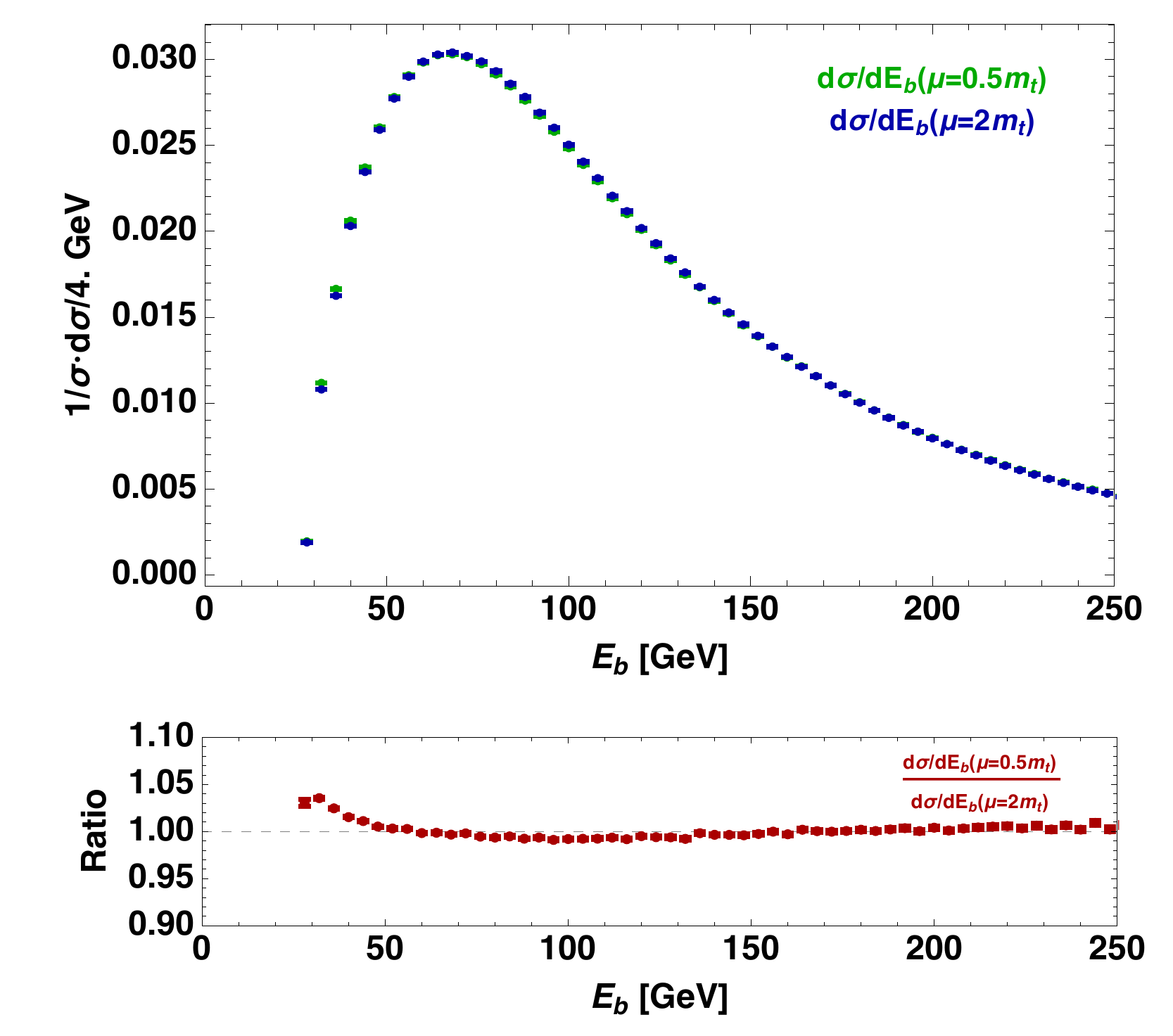}
\par\end{centering}

\caption{\label{fig:HalfTwice}
Energy spectrum under the selection criteria in Eq.~\eqref{eq:cut1} for the process Eq.~(\ref{eq:process}) at full NLO for $\mu_{F}=\mu_{R}=m_{t}/2$ (green) and $\mu_{F}=\mu_{R}=2m_{t}$ (blue) for $m_{t}=171\hbox{ GeV}$.}
\end{figure}

The prediction for $\hat{E}_{b-tag}$ as a function of the top quark pole mass is exhibited in Figure~\ref{fig:Predicted-energy-peak-jet} by the gray band.
The solid line represents the prediction for the central
scale choice, $i.e.$ $\mu_{F}=\mu_{R}=m_{t}$, and our result  for the interval shown in the figure is
\bea
\hat{E}_{b-tag}=0.55\cdot m_{t}-25\mbox{ GeV}\,,
\eea 
with RMS deviation $\delta_{rms}=0.18$ GeV.
%
The thickness of the band expresses the theory
uncertainty which reflects the difference around the peak region between the predicted energy spectra for different choices of $\mu_{R}$ and $\mu_{F}$. An example of such spectra for different choices of the scales is given in Figure~\ref{fig:HalfTwice}. 
{To better display the size of the scale sensitivity, in the bottom panel we provide the ratio of the differential energy spectrum from the process Eq.~\eqref{eq:process} at full NLO with $\mu_{F}=\mu_{R}=m_{t}/2$ to that with $\mu_{F}=\mu_{R}=2m_{t}$.}
Picking an infinitely well measured value of $\hat{E}_{b-tag}$ results in a theory uncertainty on $m_{t}$ well below $\pm1\mbox{ GeV}$.
Assuming an experimental error of $\sim1\%$ on the measured energy peak in the jet energy distribution, which is close to what was found in Ref.~\cite{CMS-PAS-TOP-15-002}, and following Ref.~\cite{Frixione:2014jk} for the definition of the errors, we find that the uncertainty on the top quark mass $\delta m_t$ is 
\bea
\delta m_t =\pm ( 1.2 \hbox{(exp)} + 0.6 \hbox{(th)} ) \hbox{ GeV}\,.
\eea
This shows that our analysis of the energy spectrum around the peak region can determine a top quark mass with $\mathcal{O}(1\gev)$ error, which is one of the main results of our work and demonstrates the utility of the energy spectrum analysis for precision mass measurements.

To test the robustness of our prediction several variations of the calculations have been executed.
Our first test is to set the central values of $\mu_R$ and $\mu_F$ at $H_T/2$.
Here $H_T$ is defined as the scalar sum of transverse momenta of visible particles, and therefore, by construction the scale values dynamically change event-by-event, as opposed to the constant $m_t$ for the baseline scale choice. 
The result for $\mu_R=\mu_F=H_{T}/2$ is reported by the
dashed line in Figure~\ref{fig:Predicted-energy-peak-jet}, which is well within the band determined from scale variation, hence corroborating the solidity of our estimate of theory uncertainty. 
Variations of the clustering algorithm have been performed as well and we found that using the $k_{T}$ algorithm~\cite{Catani:1993hr} gives insignificant differences in the predictions. 
We also examine several potential sources of the theory uncertainty in the top quark mass determination such as different definitions of selection cuts, different fitting ranges for the peak extraction, and different choices of the jet size parameter $R$. 
All these studies strengthen the conclusion that our estimate of the theory uncertainty is sound.

In reality, the transition from partons to the hadrons measured in the experiments involves more complicated dynamics than what is described by perturbative NLO corrections.
A couple of solutions are possible to improve this situation and make closer contact with experimentally measurable quantities. 
We can make use of a calculation matching NLO matrix elements and parton showers such as that of Ref.~\cite{Campbell:2014fr}, or describe directly the formation of hadrons, for example, in the fragmentation function formalism~\cite{Mele:1990cw}. 
In the next section, we will pursue the latter approach in detail, while we leave for the future a full study of the prediction of a matched calculation.
Nevertheless, here we wish to estimate the theory uncertainties carried by the parton shower. 
To this end we present results obtained from \noun{MadGraph5\_aMC@NLO} interfaced to \noun{Pythia}8, concentrating on assessing the variability of the results obtained upon changes in parameters of the parton shower. 

We compute the spectrum $d\sigma/dE_{b-tag}$ according to the same definition described earlier, preparing LO parton-level events for the process in Eq.~\eqref{eq:process} with \noun{MadGraph5\_aMC@NLO}~\cite{Alwall:2014bq} and subsequently feeding them to \noun{Pythia 8.210} for parton showering~\cite{Sjostrand:2014rr}.
We remark that the latter Monte Carlo package has been tuned to data to give a better description on heavy quark fragmentation~\cite{Skands:2014fj} and has been investigated, by the ATLAS collaboration, for the sensitivity to its parton shower parameters in the context of $t\bar{t}$ production at the LHC~\cite{ATL-PHYS-PUB-2015-007}.
Moreover, the description of extra radiation in the process Eq.~\eqref{eq:process} has been studied in detail in \noun{Pythia8}~\cite{Corke:2010qy}.
All these considerations suggest that \noun{Pythia }8 should provide a solid framework to study showering effects on our observables.

We have studied the sensitivity of $d\sigma/dE_{b-tag}$ and of $\hat{E}_{b-tag}$ to changes of the strong coupling constant used in the time-like and space-like shower in \noun{Pythia 8.210}, which probes the sensitivity of our observable to the overall amount of radiation added to the LO process, as well as the sensitivity to changes of the several cut-off scales and damping scheme of the space-like shower (Pythia parameters SpaceShower:pTdampMatch, SpaceShower:pTmaxMatch, and SpaceShower:pTmin), which probes the sensitivity of our observable to the kinematics of the shower. 
We remark that we have retained the defaults matching scheme of the time-like shower because for resonances decay, such as the top quark decay under consideration, \noun{Pythia8}
performs a matching \cite{Norrbin:2001jw,Corke:2010qy} between a matrix element description of top decay with extra jets and parton shower that is expected to be comparable in accuracy to a matched NLO calculation such as that of \noun{powheg}~\cite{Campbell:2014fr}.
Therefore, the study of the sensitivity to the value of $\alpha_{s}$ in the time-like shower is expected to give a fair estimate of the showering uncertainty for the description of radiation in the decay. 

The sensitivity to changes of parameters are quantified through the following log-derivatives with respect to the parameters listed above
\bea \label{sensitivity}
\Delta_{X}\equiv\frac{X}{\hat{E}_{b-tag}}\frac{\partial\hat{E}_{b-tag}}{\partial X}.
\eea
It turns out that the largest one is the sensitivity to $\alpha_{s}$ in the time-like shower, which is $\Delta_{\alpha_{s}}\simeq-0.05$. 
Given the importance of the description of radiation in the decay, this result is completely expected, and the sign is fully understood from the discussion around Eq.~\eqref{eq:NLOshiftJet}. 
More quantitatively, this result implies that the extracted top quark mass is shifted by more than 1 GeV if a variation greater than 12\% in $\alpha_{s}$ is considered. 
Compared to standard procedure for the variation of $\alpha_{s}$ in tuning of parton showers~\cite{Skands:2014fj} and considering the possibility to further improve these tunings in future updates of Ref.~\cite{ATL-PHYS-PUB-2015-007}, the uncertainties that arise from the physics of parton shower appear subdominant to those considered above from scale sensitivity in the fixed order calculation. 
This means that, to a very good extent, for the observable that we are proposing, the so-called ``Monte Carlo mass'' \cite{Hoang:2014la,Hoang:2008ef} coincides with the pole mass.
Although a full study of our observable in a calculation matching NLO matrix elements with parton shower is needed to quote a conclusive number, the study of the sensitivity to the parton shower parameters suggests the difference between the two masses is well below the GeV.

\section{$B$-hadrons results\label{sec:B-hadrons-results}}

Next, we discuss an extension of the energy-peak method for top quark mass measurement to $B$-hadrons. 
The correlation between the top quark mass and the $b$ quark energy from the top decay, as we have seen thus far, is largely inherited by the $b$-jet that emerges from the transition of the $b$ quark into hadrons.
Looking into the $b$-jet, we can imagine to use a single hadron in the $b$-jet (or, at least in principle, another subset of the jet constituents) and correlate its energy spectrum to the top quark mass. 
The use of jet constituents allows us to avoid experimental systematic uncertainties such as the jet energy scale, and potentially explore other systematic uncertainties, most likely theoretical ones associated to the description of hadronization. 
On top of providing an extra handle to understand the systematic uncertainties of the global set of top quark mass measurements, the use of sub-sets of the $b$-jet constituents enables us to devise further observables that can be correlated to the top quark mass (as also discussed in Ref.~\cite{Kharchilava:2000yk}). 
For instance, restricting to single hadron observables, {we can imagine to measure the peak of the energy of the tracks reconstructing a $B$-hadron,} the peak of the $B$-hadron Lorentz boosts, or the peak of its average life-time in the laboratory frame, all of which are quantities expected to inherit part of the good properties we have observed from $b$-jets spectra in the previous section. 
%
%

\begin{figure}
\begin{centering}
\includegraphics[width=0.99\columnwidth]{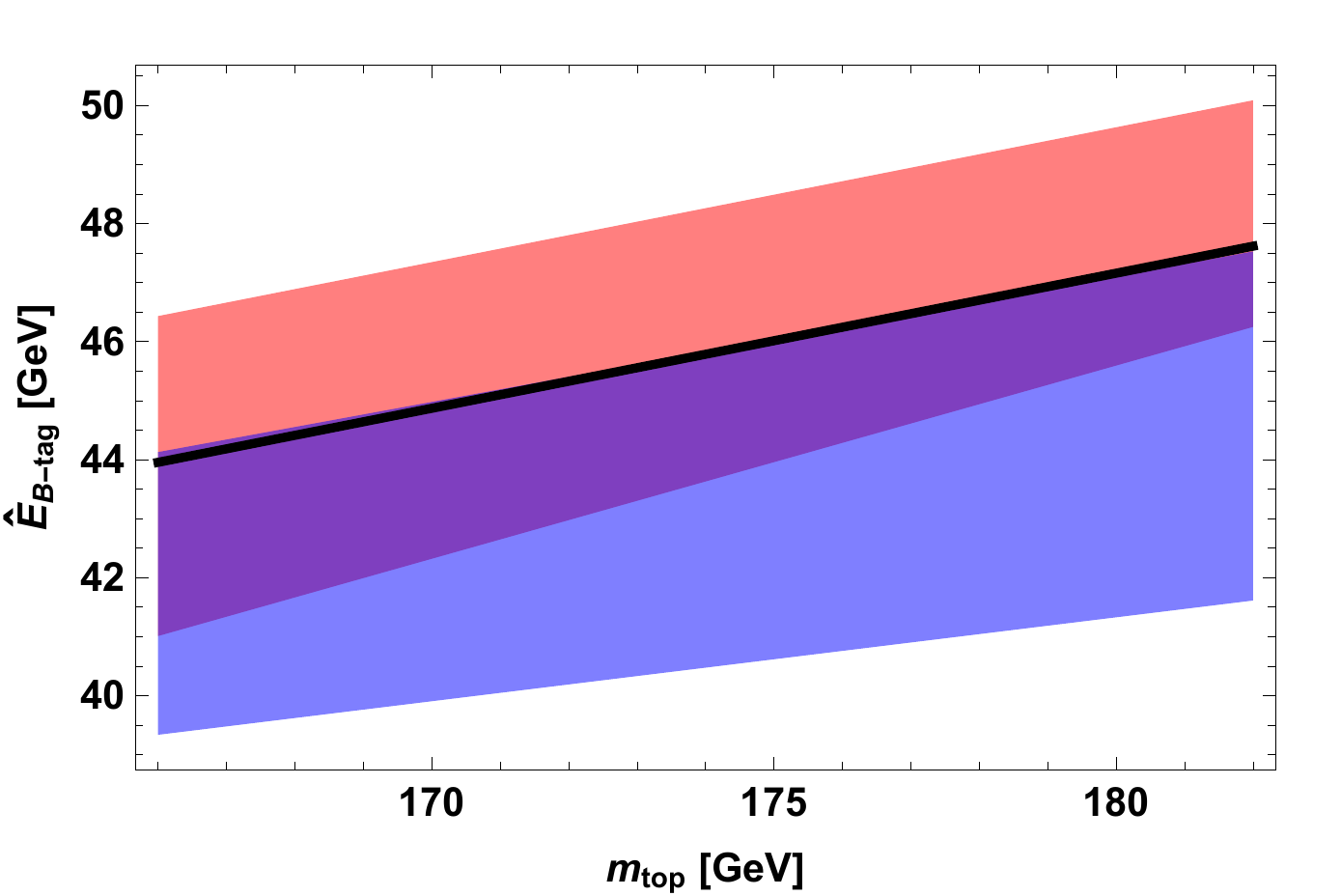}
\par\end{centering}

\caption{\label{fig:Predicted-energy-B-hadron} Predicted energy peak of the $B$-hadron spectrum as a function of $m_{t}$, as obtained from fitting the part of the spectrum where $25\mbox{ GeV}<E_{B-hadron}<100\mbox{ GeV}$ under the selection criteria in Eq.~\eqref{eq:cut1}. 
The blue and the red bands are the LO and NLO scale variations, respectively. 
As explained in the text the LO one is dominated by sensitivity to the fragmentation scale.}
\end{figure}

In order to assess the potential of the use of $B$-hadron spectra for the top quark mass measurement, we investigate the accuracy on $m_{t}$ that can be achieved from single $B$-hadrons energy spectra. 
We describe the formation of $B$-hadrons at NLO in the fragmentation functions formalism~\cite{Mele:1990cw}, performing a calculation largely analogous to that of Ref.~\cite{Biswas:2010vq} with inputs from \cite{Corcella:2005tg,Corcella:2010jl} for the parameterizations~\cite{Kartvelishvili:1977pi,Colangelo:1992kh} of the fragmentation functions. 
With this method we obtain energy spectra for $B$-hadrons for each choice of the renormalization, factorization, and fragmentation scales. 
As in the previous section, we set them to be equal and spanning the range $[0.5m_{t},2m_{t}]$. 
As the energy of the single $B$-hadron is always lower than that of the associated $b$-jet and its spectrum may also vary considerably with respect to that of $b$-jets, we have re-evaluated the details of the fitting procedure necessary to identify the maximum of the distribution. 
We find that the same fitting function applied to the case of $b$-jets is still successful, but the fit needs to be carried out on a smaller energy range, motivating us to pick $25\mbox{ GeV}<E_{B-hadron}<100\mbox{ GeV}$.
We notice that it is particularly important to avoid the lowest energy part of the spectrum, which happens to be affected by non-perturbative corrections, hence being very sensitive to variations of the scales and to the perturbative order of the calculation. 
The prediction for the peak in the $B$-hadron energy spectrum as a function of $m_{t}$ is presented as a band in Figure~\ref{fig:Predicted-energy-B-hadron}. 
The solid line is the NLO prediction for $\mu_{F}=\mu_{R}=\mu_{fr}=m_{t}$, that is,
\bea
\hat{E}_{B-hadron}=0.33\cdot m_{t}-13\, \gev\,,
\eea
{with $\delta_{rms}=0.15\hbox{ GeV}$. The thickness of the band from scale variation yields a theoretical uncertainty  $\pm3.5\mbox{ GeV}$   on $m_{t}$}. 
These results are essentially unchanged  switching between the two parametrizations of the fragmentation functions of \cite{Kartvelishvili:1977pi} and \cite{Colangelo:1992kh}.

The comparison with the results from the analysis of $b$-jets highlights a loss of the correlation between the physical observables and the top quark mass, as expected from the fact that the $B$-hadron is only a subset of the products of the $b$ quark hadronization. 
This observation motivates the investigation of other possible subsets of the $b$-jet hadronization, aiming to increase the sensitivity of the observable to $m_{t}$ still being away from using the full jet, hence not feeling jet energy scale uncertainties and possibly reducing the theory uncertainty as well. 
On this last point we remark that comparing calculations
with the fragmentation function at LO and NLO, the sensitivity to variations of only the fragmentation scales reduces from $\pm8\mbox{ GeV}$ to $\pm4\mbox{ GeV}$, hence motivating improvement to NNLO fragmentation functions in the future.

\section{Conclusions\label{sec:Conclusions}}
In Ref.~\cite{Agashe:2013sw} a theoretical proposal for determining the top quark mass via a measurement of the position of the peak in the
energy distribution of the $b$-jets resulting from its decay was made.
In the present paper we elevate this technique to the precision frontier by computing the correlation between this energy-peak and the top quark mass at NLO in QCD. 
The results presented here reduce the theory uncertainty on the original method and are particularly useful for the recent implementation by the CMS collaboration~\cite{CMS-PAS-TOP-15-002}, which was based on LO calculations.
The remaining theory uncertainty from sensitivity of the $b$-jet energy spectrum to relevant parameters has been investigated in detail
 by studying several variations of the calculation, including that of jet radii and dependence of the factorization and renormalization scales on the momenta of the top decay products. 
The uncertainties from multiple emissions, to be described in the parton shower approximation, have been estimated as well.
The overall result is that the $b$-jet energy-peak method can be used to extract the top quark pole mass with an uncertainty $\pm (1.2 (\hbox{exp}) + 0.6(\hbox{th})) \hbox{  GeV}$. 
Here we have assumed large statistics so that the experimental uncertainty, which dominates the theory one, comes primarily from the jet energy scale uncertainty.
The method that we propose can therefore contribute significantly to the precision determination of the top quark mass at the LHC and especially to the understanding of the systematic uncertainties on the measurements. 
This value 
added by 
%
%
%
the technique is manifest partly from the
above-quoted error being comparable to that from the 
%
%
existing methods.
Moreover, the underlying idea of the measurement has certain merits with respect to other methods. 
Firstly, our method belongs to the class of mass measurement methods based on relating the top quark mass to a kinematical observable. 
In this approach, the correlation between the top quark mass and the chosen observable can be studied in QCD including higher order corrections, with uncertainties that can be quantified in reliable ways, as we did in this work.  
This implies that our method, as well as all the others based on a kinematical observable, does not suffer from some difficulties that are instead typical of ``matrix element methods", {\it i.e.} based on the measurement of {\em all} the observed final states, thus
requiring knowledge of the {\it complete} matrix element of the SM for top quark production and decay ({\it e.g.} \cite{D0-Collaboration:2015rm}).

Furthermore, in the landscape of recent proposals based on kinematical observables, our method emerges for  having reduced sensitivity to details of the production mechanism of the top quark. 
This insensitivity arises because the observable we used, namely the  peak of the energy distribution, has been shown~\cite{Agashe:2013sw} to enjoy a certain degree of invariance under boost distributions of the top quark.
This characteristic  is underscored by the formula for the top quark mass in terms of the observable in our method, at leading order for an inclusive top decay, being known simply and analytically from the prime principles of the kinematics of decay {\em only}, regardless of the production mechanism of the $t\bar{t}$ pair (as long as the resulting top quarks are unpolarized). 
On the contrary, for the other kinematic observables,\footnote{A notable exception being the CMS endpoint analysis \cite{CMS-PAS-TOP-11-027}.} already at leading order in perturbation theory and in absence of cuts, it is necessary to carry out a {\em detailed} calculation to 
%
%
determine
the correlation with the top quark mass.
This means that a {\em specific} matrix element, $e.g.$ that of the SM, has to be employed in addition to knowledge of the collider beam energy and parton distribution functions. 
The resulting  advantage of our method over others is hence clear -- the top mass extracted using our observable is subject to smaller error from higher-order QCD effects in {\em production} and other variations of the production mechanism such as 
%
%
the 
parton distribution functions set that is used.
The above expectation is corroborated for instance by the smaller error on the measured top quark mass associated with the mis-modelling of the top $p_{T}$ spectrum that has been reported using 
the $b$-jet energy-peak method by the CMS collaboration in Ref.~\cite{CMS-PAS-TOP-15-002} in comparison to other methods, {\it e.g.} the 
$B$-hadron decay length 
%
%
method Ref.~\cite{CMS-PAS-TOP-12-030}. 
{We remark that the $B$-hadron decay length
is {\em correlated} with the energy of the $B$-hadron (in turn, with that of the parent 
bottom quark) so that (naively) one would expect
that the sensitivities to the above top quark $p_{T}$ modeling should have been  similar in the two methods. 
However, the $B$-hadron decay length analysis does not take advantage of the energy peak result in its implementation, because the actual decay length observable used in this CMS measurement did {\em not} correspond exactly to the $B$-hadron energy-peak being discussed here, thus it appears justified that the decay length measurement does not enjoy a 
reduced sensitivity to production details.}
Relatedly, the  method proposed in this work is also minimally sensitive to additional mechanism of production of top quarks from new physics, as for instance stop pair production and decay into unpolarized pairs of on-shell top quarks $pp\to \tilde{t} \tilde{t}^{*} \to t \bar{t} \chi \chi$ in supersymmetric extensions of the SM. 
Therefore, the method that we propose adds important
%
%
information to the measurements of other methods that, still being based on a kinematical observable, assume the SM production mechanism.
On the flip side, our method has sensitivity to QCD effects and new physics if they change top quark decay distributions. 
This complementarity implies that it is clearly useful to compare results obtained using our strategy with those obtained with others.

%
%
The method that we discuss also might have reduced sensitivity to {\em parton showers} (PS) 
and details of their implementation for the following reason [as implied by the discussion 
related to
%
%
%
Eq.~(\ref{sensitivity})]. 
%
%
Namely, our observable is mostly affected by {\em hard} emissions (in the decay
%
%
i.e. much less in the production), which are best described using the fixed order matrix elements. 
%
%
%
%
%
%
%
In any case, 
due to this sensitivity also coming mostly from corrections to the {\em decay}, we expect the analysis of energy peaks to carry complementary information about the effect of PS
%
%
to that obtained from other observables which instead (in general) feel it via both production {\em and} decay corrections.
Furthermore, using existing NNLO QCD~\cite{Gao:2013wk,Brucherseifer:2013mg} corrections to the top quark decay it would be possible to keep using well-defined QFT mass definition to compute our observable to even {\em higher} accuracy. 
Hence, we can minimize the contribution of inevitable further emissions and other effects not described by a fixed order calculation, possibly attaining 
%
%
a
top quark mass measurement free from the need to talk about a ``Monte Carlo mass''~\cite{Hoang:2008ef,Hoang:2014la}.
In this context of assessing the effects of multiple
emissions, it is going also to be useful to perform a study of our observable in a matched NLO-PS calculation~\cite{Campbell:2014fr}, which will be presented elsewhere.
Note, however, that the results presented in this work on the sensitivity of the extracted mass to 
%
%
variations 
of the shower parameters clearly indicate that is possible to treat this mass as the pole mass up to corrections well below the GeV. 

As extensions of the measurement methods based on energy spectra, we anticipate the use of {\em sub}-jets and single hadron properties ($cf.,$~inclusive $b$-jet above). 
The $B$-hadron energies can be in principle determined free of jet energy scale, which is the dominant source of experimental uncertainty on many top quark mass measurements (including the energy-peak of $b$-jets).
For example, this can be done using only tracker measurements of the charged decay products of a $B$-hadron. 
Therefore, pursuing a reduction of the experimental error, 
we have studied energy spectra of $B$-hadrons and showed that the top quark mass can be extracted form these energy-peaks as well, though with larger uncertainties coming from the bottom quark fragmentation functions.
This theory uncertainty on the top quark mass measurement from the $B$-hadron energy-peak can in principle be reduced using NNLO fragmentation functions which might become available in the future. 
Such an improvement might bring the methods based on $B$-hadron energy spectra (in general) to a level of accuracy similar to the current ones.

{As mentioned above, the CMS collaboration has already measured the top quark mass using $B$-hadron decay length as a ``proxy" for the {\em energy} of the $B$-hadron.
Although this Run I analysis assumed {\em SM} production, it did ameliorate the jet energy scale uncertainty along the above lines, as it used 
(mostly) information from the tracker. 
For the Run II measurements, it would then
be interesting to have the best of both worlds by pursuing a {\em combination} of these two ideas, $i.e.$ augment the $B$-hadron decay length method
by the energy-peak result so that the resulting method can have
reduced sensitivity to jet energy scale uncertainty {\em and}
modeling of top quark production.}

In summary, the results of this work demonstrate the usefulness of energy spectra (in particular, their peaks) for precision mass measurements and illustrate the possible benefit of inclusion of this type of analysis in future combined extraction of the top quark mass at the LHC.

\begin{acknowledgments}

KA is supported in part by NSF Grant No. PHY-1315155 and by the Maryland Center for Fundamental Physics. 
DK is supported by the Department of Energy under
Grant DE-SC0010296.
RF thanks the LHC TOP Working Group and  S. Amoroso,  F. Boudjema, M. Cacciari,  F. Caola, R. Chierici, G. Corcella, S. Davidson, J. Fuster, S. Frixione, A. Hoang, E. Laenen,  M. Mangano, A. Mitov, S. Moch,  M. Mulders, P. Nason, E. Re,  G. Salam, P. Silva, T. Sj\"{o}trand, P. Skands, B. Stieger, P. Uwer, M. Vos, A. Weiler, G. Zanderighi for their input and very interesting, useful comments as this work  progressed.

\end{acknowledgments}

\bibliographystyle{apsrev4-1}
%

\end{document}